\documentclass[preprint,12pt]{elsarticle}

\usepackage{amssymb}

\usepackage{hyperref}

\journal{Computer Physics Communications}

\begin{document}

\begin{frontmatter}



\title{Efficiency of linked cell algorithms}


\author{Ulrich Welling}
\ead{ulrich.welling@staff.uni-marburg.de}
\author{Guido Germano}
\ead{guido.germano@staff.uni-marburg.de}
\address{Department of Chemistry and WZMW, Philipps-University Marburg,
35032 Marburg, Germany}
\ead[url]{http://www.uni-marburg.de/fb15/ag-germano}

\begin{abstract}
The linked cell list algorithm is an essential part of molecular simulation
software, both molecular dynamics and Monte Carlo. Though it scales linearly
with the number of particles, there has been a constant interest in increasing
its efficiency, because a large part of CPU time is spent to identify the
interacting particles. Several recent publications proposed improvements to
the algorithm and investigated their efficiency by applying them to particular
setups. In this publication we develop a general method to evaluate the
efficiency of these algorithms, which is mostly independent of the parameters
of the simulation, and test it for a number of linked cell list algorithms.
We also propose a combination of linked cell reordering and interaction sorting
that shows a good efficiency for a broad range of simulation setups.
\end{abstract}


\end{frontmatter}


\section{Introduction}
\label{Introduction}
A fundamental issue in the evaluation of non-bonded interactions in molecular
simulation is to avoid the calculation of all pair distances because particles
beyond a certain cutoff radius $r_\mathrm{c}$ can be neglected
\cite{AllenTildesley1989}. The two main methods to deal with this are the
Verlet list \cite{AllenTildesley1989,Verlet1967,Chialvo1983} and the linked
cell list (LCL) \cite{AllenTildesley1989,Quentrec1973,Hockney1981}.
A further improvement uses a LCL to generate a Verlet list \cite{Mattson1999,
Petrella2003,Yao2004,Wang2007}. These methods apply to both molecular dynamics
(MD) and Monte Carlo simulations, but for convenience here we use only MD to
illustrate them.

The reason to apply more complex algorithms than a plain LCL is that the
majority of the computed pairwise distances is not within the cutoff radius.
So, despite of the linear scaling with the number of particles, there is space
for further optimizations. Verlet lists solve this accurately by taking into
account only particles within a Verlet radius $r_\mathrm{V} > r_\mathrm{c}$.
However Verlet lists have limitations for large scale simulations, as the total
memory needed by them scales as the number of particles $N$ times the average
number of neighbours. Moreover they must be updated regularily and thus the
same basic problem of determining neighbours within a cutoff needs to be
solved.

Mason \cite{Mason2005} introduced a variant of the LCL where, instead of
dividing particles into cells with cubic shape, an arbitrary lattice is set up
and particles are sorted into a container belonging to the closest lattice
point. This can be seen as a generalization of the LCL by allowing all possible
coordination polyhedra.

\begin{figure}[h]
\begin{center}
\includegraphics[width=0.7\columnwidth,clip=true]{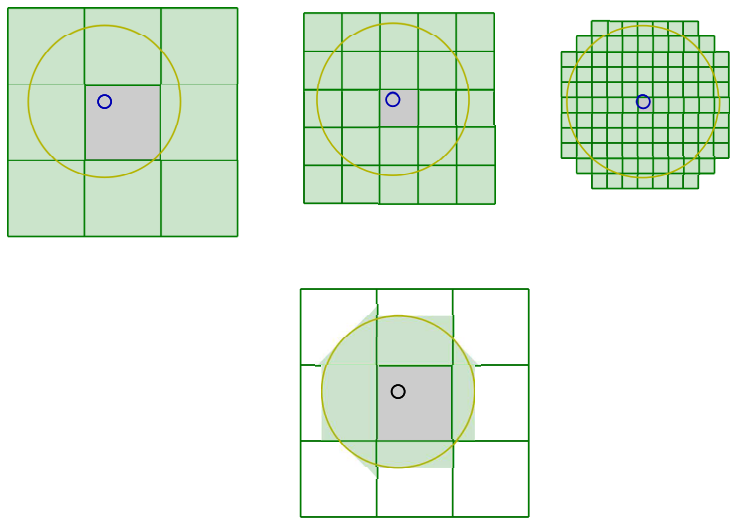}
\end{center}
\caption{\label{fig:cells_gonnet}
Reduction of the search space (grey/green) using smaller cells (top) or
Gonnet's method (bottom).}
\end{figure}

Gonnet \cite{Gonnet2007} modified the LCL to avoid the majority of unnecessary
distance calculations and thus gaining a significant speedup with respect to
the standard algorithm by sorting particles according to their projection onto
the vector that connects the cell centers; a schematic picture is given in
Fig.~\ref{fig:cells_gonnet}. However in his publication he only investigated
the use of this method for one specific case, i.e.\ the evaluation of the real
space part of the Coulomb interaction of water molecules. We wished to
investigate the performance of this algorithm for a wide area of simulation
setups, taking into account the results of other publications concerning the
optimization of linked cell algorithms.

To this aim a general method for comparing the performance of linked cell
algorithms is needed. Sutmann \cite{Sutmann2006} tested generalized LCLs for 
Lennard-Jones systems with different densities and cutoff radii. One of the
results is that the ratio between the cell side $L$ and the cutoff has an
optimal value typically for a fraction of integer numbers, for example $1:1$ or
$1:2$, which can be deduced by geometrical considerations, as the sphere that
is built by the cutoff radius fits perfectly a cubic cell, or an array thereof,
only in these cases. However the cutoff radius and the density of the system
are connected by another parameter, which is the average number of neighbours
each particle interacts with, i.e.\ the particles that reside within the cutoff
radius around each particle. This single parameter is the most fundamental one
for judging the performance of the different linked cell algorithms. As all
linked cell algorithms scale more or less linearly with increasing number of
particles, we can assume that an algorithm that works fastest for a given
average number of interactions will do so regardless of the size of the
simulation.

Apart from purely algorithmic considerations, there are publications
addressing an effect that can be summarized as cache-optimized memory access
\cite{Yao2004,Kadau2006,Meloni2007}. The idea is to rearrange the particle
information in order to ensure that the data representing particles which are
close to each other in the simulated system is located as close as possible in
memory, in order to achieve cache hits within the hierarchical memory typical
of contemporary computers. Computer memory can be seen as a one-dimensional
array, but typical molecular simulations are done in three spatial dimensions,
so a perfect mapping is impossible. One early approach was to sort the
particles into slabs along one of the three spatial axes, preferably the
one corresponding to the largest box side \cite{Hockney1981,Yao2004}.
Meloni et al.\ \cite{Meloni2007} picked up a principle that had been used in
multi-billion particle simulations \cite{Kadau2006} and demonstrated the gain
in efficiency by reordering particles in memory by their linked cell index,
which is referred to as linked cell reordering (LCR). A further improvement of
data locality can be achieved doing this by means of the space-filling Peano or
Hilbert curve \cite{Anderson2008}. However rearranging particles in memory is
a hardware specific optimization that will not necessarily be an advantage on
all kinds of computers. Specifically, approaches that take advantage of
coprocessors like graphic processing units \cite{Anderson2008,Liu2008} or
cell processors might need different adjustments.

To investigate the efficiency of the different linked cell algorithms we set up
a small scalar MD program in C++ for the Lennard-Jones fluid and implemented
the different linked cell algorithms, taking into account considerations of
several publications concerning aspects for an efficient implementation.

\section{Minimum image convention}

As pointed out by P\"utz and Kolb \cite{Putz1998} as well as by Heinz and
Huenenberger \cite{Heinz2004}, the minimum image problem needs to be adressed
on the linked cell level rather than for every particle pair. In the approach
given in the book of Allen and Tildesley \cite{AllenTildesley1989} the minimum
image convention is computed for every calculated distance. This is done either
with modulo operations or conditional tests, both of which are expensive in
computation time. A better way is to compute a distance vector for an evaluated
cell pair and to add it to the distance vector of molecules within these cells.
This way the periodic images are resolved by three additions for any calculated
distance. In this work we used the method introduced by Rapaport
\cite{Rapaport1988} and used by P\"utz and Kolb \cite{Putz1998}, which uses
copies of the linked cells at the borders of the unit box with translated
coordinates. This way the evaluation of the pairwise forces does not need any
additional operation. This is similar to domain decomposition, where the
particles in the cells at the boundary of a domain are ``exported'',
i.e.\ copied, to the neighbour domains for the calculation of the forces
\cite{Rapaport1988,Plimpton1995,Wilson1997,Heffelfinger2000,Shaw2005,Wu2005}.
Only interactions with the positive half shell of surrounding cells are
computed, so Newton's third law is exploited to avoid the calculation of an
interaction twice. The forces on the particles in the copied region are
reassigned to the original particles. This reduces the effort of computing the
minimum image convention from being proportional to the number of evaluated
pair distances to the number of particles within the border region of the unit
box. Gonnet \cite{Gonnet2007} used the variant with an additional vector to
resolve periodicity, thus making distance evaluations a little more expensive
than necessary. It is important to avoid these superfluous computations, as it
is possible that a performance gain of a different linked cell algorithm could
be caused just by reducing the number of minimum image calculations, as can be
seen in Mason's publication \cite{Mason2005}. He explicitly writes that in his
approach minimum images are computed per particle pair and that this
computation is time consuming.

\section{Linked cell variants}

In our test program we implemented several variants of linked cell
algorithms. As basis we use a generalized LCR with an arbitrary relation
between the cutoff radius and the cell side. During the initialization the
program computes an interaction matrix of the linked cells given in relative
coordinates as proposed by Mattson \cite{Mattson1999}.

The LCR algorithm as given by Kadau \cite{Kadau2006} and Meloni et al.\
\cite{Meloni2007} consists in computing at every time step the scalar cell
index $i$ of each particle from its coordinates and then to sort the particles
by $i$. The latter is computed e.g.\ as $i = i_z + i_y n_z + i_x n_y n_z$,
where $i_\alpha = 0, \dots, n_\alpha-1$ are the Cartesian cell grid indices.
An improvement computes $i$ through the space-filling Peano or Hilbert curve
\cite{Anderson2008}, which results in better data locality. The sorting can
be achieved by a hash sort which takes $O(N)$ operations, as the numbers of
particles with the same cell index can be counted, and in a second loop all
particles can be moved directly to a position according to their cell index.
An advantage of the reordering method is that within one time step only a few
particles change their cell, so the effort of rearranging the particles is
small. Interestingly a quicksort \cite{Knuth1973} that exploits pre-existing
order can be applied for sorting; in simulations with slow moving particles
this is slightly faster than the direct way. We also generalized the LCR method
for arbitrary fractions $m/n,\ m,n \in \mathbb{N},$ of the cutoff and the cell
side. For this a list of relative cell coordinates that are within the cutoff
radius around a central cell is generated and used to define the interacting
cell pairs during the force evaluation. This list is reduced by a factor two
exploiting Newton's third law as mentioned in the previous section.

On top of this we implemented Gonnet's interaction sorting method
\cite{Gonnet2007}. As it evaluates cell pairs, it can be used with all the
linked cell algorithms, both with list and sorted variants and with an
arbitrary cell-side to cutoff ratio. The full details are discussed in
Gonnet's publication; the principle is that for each pair of cells all the
particles are projected onto the vector that connects the cell centers. The
projection can be simplified in the special cases where one or two components
of the connecting vector are zero, which happens when the cells have one or two
identical Cartesian grid indices. However, the necessary additional conditional
evaluations are slower than the saved multiplications by zero, and the code is
more complex.

Afterwards the particles of both cells are sorted by the projected distance.
Since this value is smaller than the distance itself, if the former is larger
than the cutoff the latter will be larger too and this pair of particles does
not need to be evaluated. So instead of computing all pairs between the two
cells, which would be an operation of order $O(N_1 N_2)$, all the particles of
both cells are projected and sorted, with an effort of $O(N_1 + N_2 + N_1
\log N_1 + N_2 \log N_2)$. The task to be done here is sorting a small set of
particles as fast as possible. Gonnet used a typical implementation of the
quicksort method that stops at a certain array size, usually smaller than 10,
and handles the remaining subarray with an $O(N^2)$ method like insertion sort.
However, for small sets the best solution are optimal sorting networks
\cite{Knuth1973}. They result in the minimum number of comparisons and swaps,
close to the theoretical limit of $N \log_2 N!$ operations. Thus in our
implementation we resorted to a combination of the quicksort algorithm and,
for each subarray size of 16 and smaller, a suited optimal sorting network.
Gonnet's result happened to be faster than the generalized LC algorithm with
$L = r_\mathrm{c}/2$ though his sorting method was not optimal because in both
cases he computed the minimum image convention in a fashion that was not
optimal either. Speeding up the latter, it becomes essential to improve the
sorting too in order to remain competitive with respect to the generalized LC
algorithm without interaction sorting.

\section{Computational details}

To evaluate the efficiency we did a set of simulations with changing average
number of interactions per particle. A typical setup of the Lennard-Jones fluid
with a density $\rho = 0.72$ and a cutoff radius $r_\mathrm{c} = 2.5$ has an
average number of interactions of $(4/3)\pi r_\mathrm{c}^3\rho = 55$. Gonnet
tested his modification of the linked cell algorithm for the real space part of
the Coulomb interaction in water with $r_\mathrm{c} = 10\,$nm, resulting in an
average number of interactions of 150. We are not aware of MD simulations with
a considerably larger number of interactions. For each average number of
interactions per particle we obtain the average time per integration step and
particle.

We set up a small scalar MD program for the Lennard-Jones fluid using the
velocity Verlet integrator \cite{Swope1982}. Periodic boundary conditions are
applied with the minimum image convention computed as described before. For the
tests we generated particles on a lattice with a density derived from the
required number of interacting particles. For simplicity our unit box and
therefore also its subcells were cubic. The velocities were randomized and the
system was equilibrated for 20\,000 steps of $\Delta t = 0.001$ at a
temperature $T = 5$ using velocity scaling, followed by 4000 steps of time
profiling in the microcanonical ensemble. We used the GNU C++ compiler version
4.3.2 to produce 64\,bit binaries with double precision on an Intel Core2 Quad
running at 1.6\,GHz.

\section{Results and discussion}

\begin{figure}[h]
\begin{center}
\includegraphics[angle=270,width=0.9\columnwidth,clip=true]{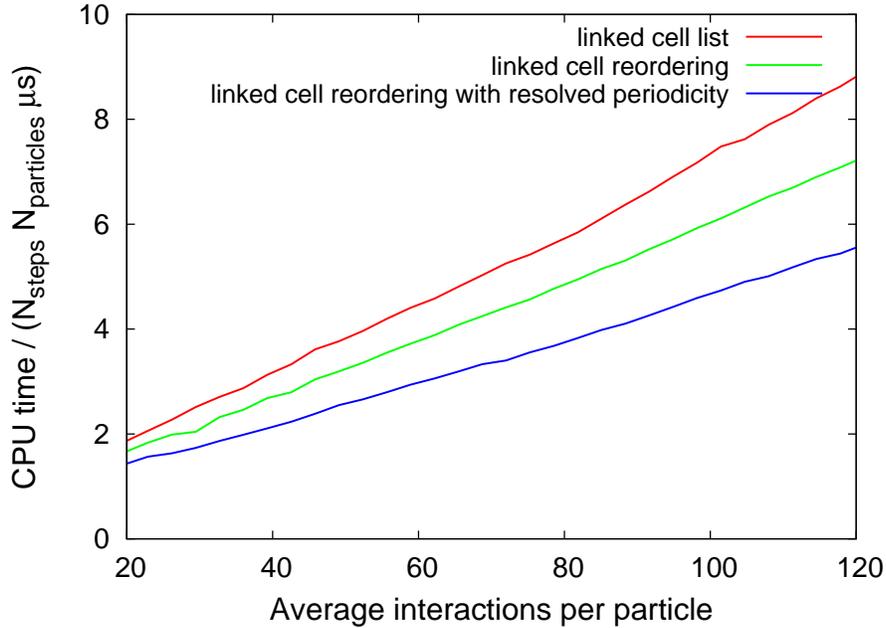}
\end{center}
\caption{\label{fig:linkedcells}
Performance of linked cell variants.}
\end{figure}

First of all we compared the LCL, the LCR with a minimum image check for every
calculated distance, and the LCR with resolved periodicity as described above 
(Fig.~\ref{fig:linkedcells}). The impact on performance of avoiding the
evaluation of the minimum image convention is higher than changing from the
list to reordering.

\begin{figure}[h]
\begin{center}
\includegraphics[angle=270,width=0.9\columnwidth,clip=true]{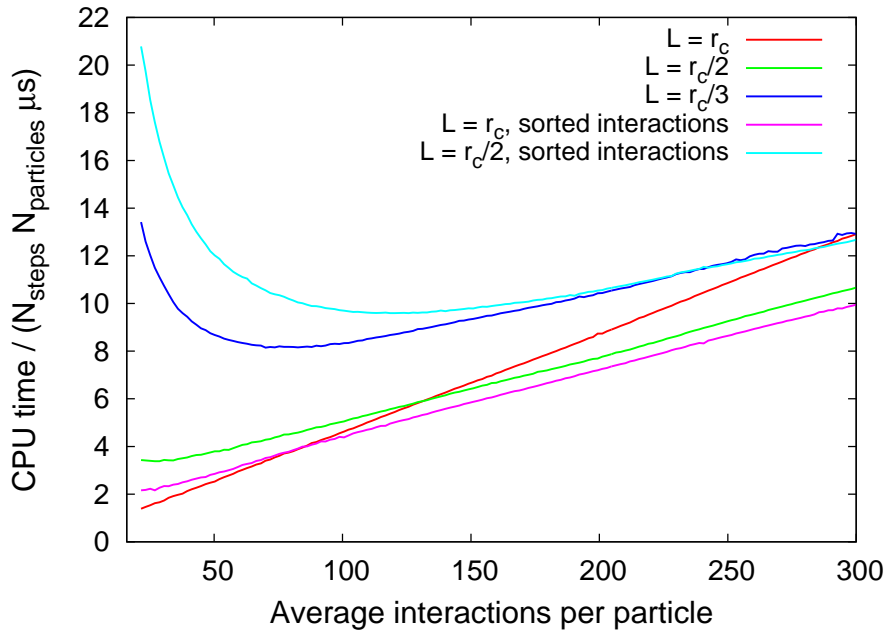}
\end{center}
\caption{\label{fig:perfectfit}
Performance of the generalized LCR with and without sorting and resolved
periodicity, at different ratios between the cell side $L$ and the cutoff
radius $r_\mathrm{c}$.}
\end{figure}

Knowing this we compared Gonnet's extention with the generalized LCR variants
to see which algorithm performs fastest for any given average number of
interactions (Fig.~\ref{fig:perfectfit}). Gonnet's algorithm does perform
faster than all the generalized LCR variants. The break-even between the LCR
with a cell side equal to the cutoff radius and Gonnet's modified method is
approximately at an average number of interactions equal to 82. This
corresponds to an average number of 19 particles per linked cell.

\begin{figure}[h]
\begin{center}
\includegraphics[angle=270,width=0.9\columnwidth,clip=true]{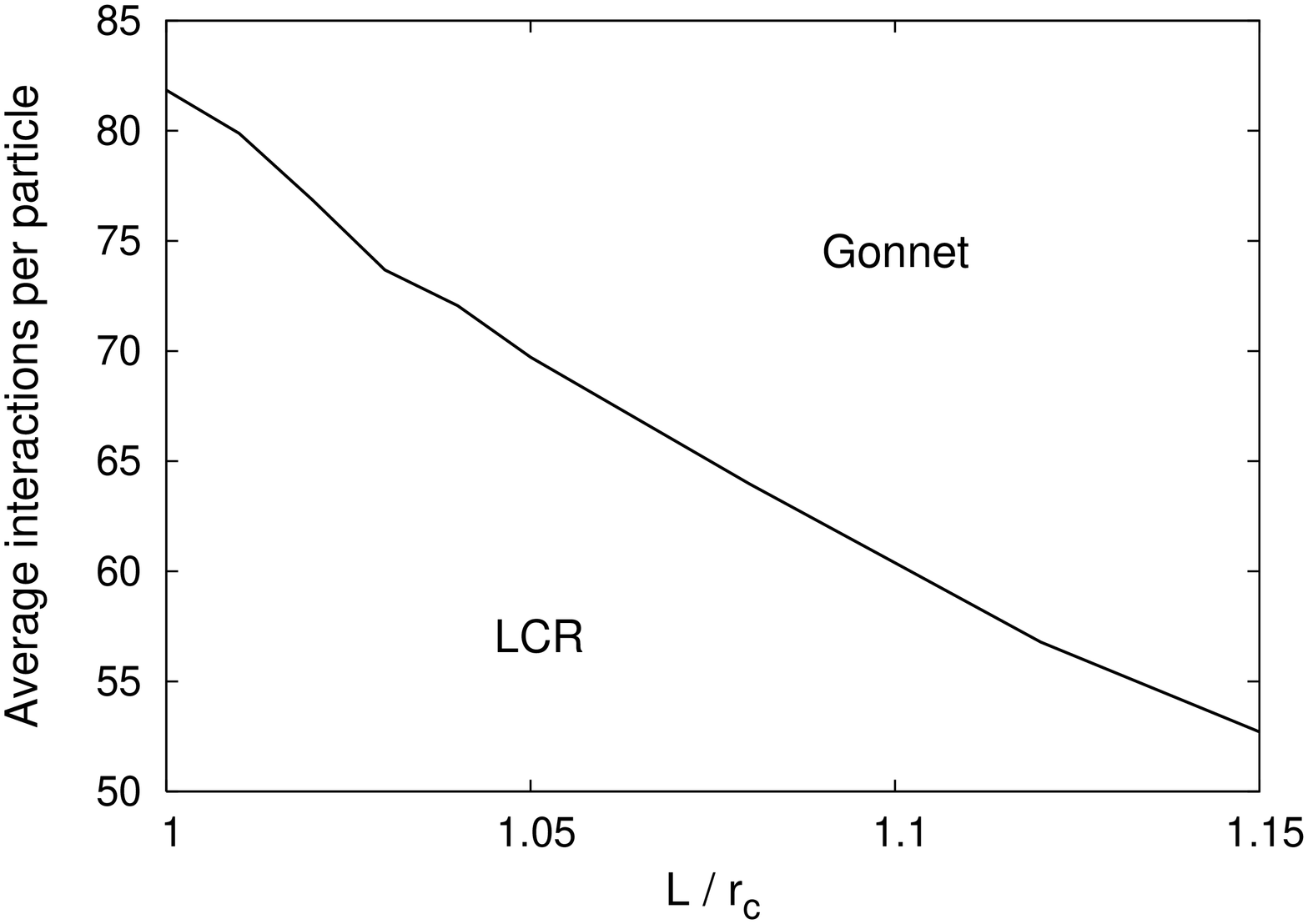}
\end{center}
\caption{\label{fig:crossover}
Performance crossover for simulations with cell sides $L$ slightly larger than
the cutoff radius $r_c$. Below LCR is faster, above the Gonnet method is
faster.}
\end{figure}

\begin{figure}[h]
\begin{center}
\includegraphics[angle=270,width=0.9\columnwidth,clip=true]{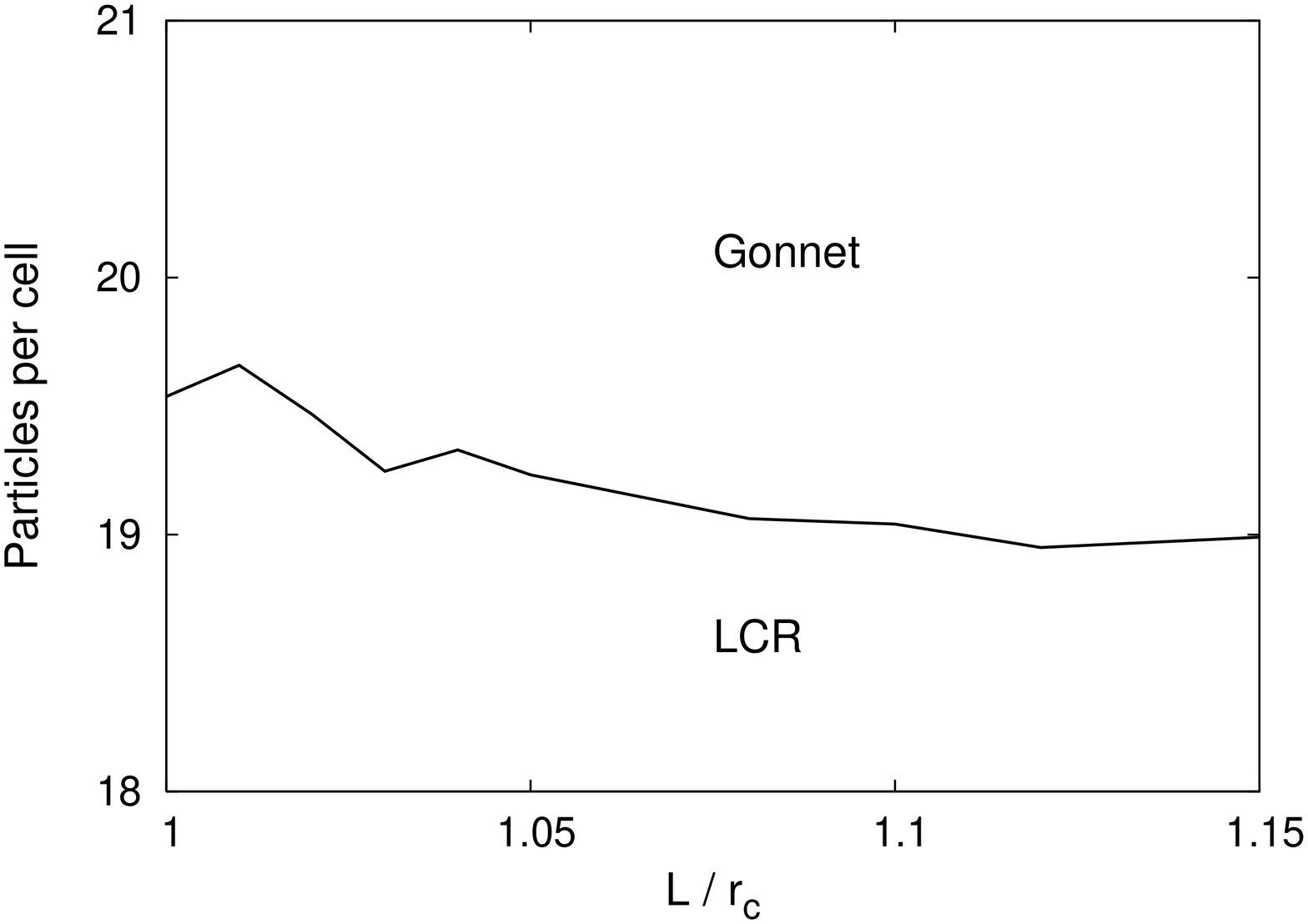}
\end{center}
\caption{\label{fig:crossover_cells}
Performance crossover for simulations with cell sides $L$ slightly larger than
the cutoff radius $r_c$. Below LCR is faster, above the Gonnet method is
faster.}
\end{figure}

Usual MD simulations do not set the side of the unit box as an integer multiple
of the cutoff, so that the real size of the linked cells is generally slightly
larger than the cutoff radius. We plotted the position of the crossover as a
function of the ratio between the side of the linked cells and the cutoff
(Fig.~\ref{fig:crossover}). Gonnet's approach tolerates larger linked cell
sides much better than the standard linked cell algorithm. If we put the
average number of particles per linked cell on the $y$-axis rather than the
average interaction count per particle, we see that the crossover stays between
19 and 20 particles per linked cell (Fig.~\ref{fig:crossover_cells}). This is
almost a constant that is pretty much unaffected by all the simulation details.
A MD software could automatically choose between the standard linked cell
implementation and Gonnet's method just using the number of particles per
linked cell as a criterium. This value depends on the compiler and a lot of
hardware specific parameters like cache sizes and the CPU architecture; thus it
must be determined separately for each system.

Even for a Lennard-Jones fluid simulation with $r_\mathrm{c} = 2.5$ and
$\rho = 1.0$ yielding an average interaction count of 65, Gonnet's method will
be faster if the ratio $L/r_\mathrm{c}$ is larger than 1.08, which can easily
happen for small-sized simulations.

If the minimum image convention is resolved in a more expensive way than we
did in our test, Gonnet's method is faster for even a smaller number of
interactions or particles per linked cell.

\section{Summary}

We investigated Gonnet's variant of the linked cell algorithm. We modified it
in two ways, using linked cell reordering rather than a plain linked list and
changing the sort method to a combination of quicksort and optimal sorting
networks that are particularly suited for small arrays. Moreover we used the
fastest known approach for the computation of the minimum image convention,
which resembles the scheme of domain decomposition parallelism. Then we
evaluated the performance of the modified Gonnet approach in comparison with
the known linked cell algorithm variants, and introduced a general way to
compare the performance of these algorithms independently of the simulation
parameters. On our test systems the reordering version of the linked cell
algorithm performed better than the list variant. This basically reproduces the
work by Meloni et al.\ \cite{Meloni2007}, however we investigated the behaviour
for a broad range of possible simulation setups. Gonnet's modified variant
always outperformes the generalized linked cell variants in our tests and is
only slower than the linked cell algorithm for less than roughly 19 particles
per linked cell. Our results show that even for Lennard-Jones fluids cases
exist where the modified Gonnet variant of the linked cell algorithm is faster
than the standard linked cell algorithm. These results enable molecular
dynamics software to choose the faster method based on simple statistics that
can be computed easily during a simulation.

\section*{Acknowledgments}
We thank Pedro Gonnet for providing a reference implementation of his
algorithm.





\bibliographystyle{elsarticle-num}
\bibliography{paper}







\end{document}